# Determination of thorium, uranium, and potassium elemental concentrations in surface soils in Cyprus

Michalis Tzortzis and Haralabos Tsertos[*]

*Department of Physics, University of Cyprus, P.O. Box 20537, 1678 Nicosia, Cyprus.*

(Revised version: 8/03/2004)

## Abstract

A comprehensive study was conducted to determine thorium, uranium and potassium elemental concentrations in surface soils throughout the accessible area of Cyprus using high-resolution γ–ray spectrometry. A total of 115 soil samples was collected from all over the bedrock surface of the island based on the different lithological units of the study area. The soil samples were air-dried, sieved through a fine mesh, sealed in 1000-*mL* plastic Marinelli beakers, and measured in the laboratory in terms of their gamma radioactivity for a counting time of 18 *hours* each. From the measured γ–ray spectra, elemental concentrations were determined for thorium (range from $2.5 \times 10^{-3}$ to 9.8 $\mu g\ g^{-1}$), uranium (from $8.1 \times 10^{-4}$ to 3.2 $\mu g\ g^{-1}$) and potassium (from $1.3 \times 10^{-4}$ to 1.9 %). The arithmetic mean values (A.M. ± S.D.) calculated from all samples are: (1.2 ± 1.7) $\mu g\ g^{-1}$, (0.6 ± 0.7) $\mu g\ g^{-1}$, and (0.4 ± 0.3) %, for thorium, uranium and potassium, respectively, which are by a factor of three to six lower than the world average values of 7.4 $\mu g\ g^{-1}$ (Th), 2.8 $\mu g\ g^{-1}$ (U) and 1.3 % (K) derived from all data available worldwide. The best-fitting relation between the concentrations of Th and K versus U, and also of K versus Th, is essentially of linear type with a correlation coefficient of 0.93, 0.84, and 0.90, respectively. The *Th/U*, *K/U,* and *K/Th* ratios (slopes) extracted are equal to 2.0, 2.8×10$^3$, and 1.4×10$^3$, respectively.

---

[*] **Corresponding author.** E-mail address: *tsertos@ucy.ac.cy*, Fax: +357-22892821.

# 1. Introduction

Gamma radiation emitted from naturally occurring radioisotopes, such as $^{40}$K and the radionuclides from the $^{232}$Th and $^{238}$U series and their decay products (also called terrestrial background radiation), which exist at trace levels in all ground formations, represents the main external source of irradiation to the human body. More specifically, natural environmental radioactivity and the associated external exposure due to gamma radiation depend primarily on the geological and geographical conditions, and appear at different levels in the soils of each region in the world (UNSCEAR 2000 Report, and further references cited therein). The specific levels of terrestrial environmental radiation are related to the geological composition of each lithologically separated area, and to the content in thorium (Th), uranium (U) and potassium (K) of the rock from which the soils originate in each area. In terms of natural radioactivity, it is well known, for instance, that igneous rocks of granitic composition are strongly enriched in Th and U (on an average 15 $\mu g\ g^{-1}$ of Th and 5 $\mu g\ g^{-1}$ of U), compared to rocks of basaltic or ultramafic composition (< 1 $\mu g\ g^{-1}$ of U) (Faure, 1986; Mènager et al., 1993). For that reason, higher radiation levels are associated with igneous rocks and lower levels with sedimentary rocks. There are exceptions, however, as some shales and phosphate rocks have relatively high content of those radionuclides (UNSCEAR 2000 Report).

From a geological point of view, the island of Cyprus, which is located in the eastern basin of the Mediterranean Sea and extends to an area of about 9,300 $km^2$, contains one of the best-preserved and most intensively studied ophiolite complexes in the world, which is known as the Troodos Massif (Moores and Vine, 1971; Robinson and Malpas, 1998). It consists of basic and ultrabasic pillow lavas, fringed by andesitic-sheeted dykes, while the central part of the ophiolite consists of basic and ultrabasic plutonic rocks (gabbros, peridotites, dunites and serpentinised harzburgites). The highly tectonised and fractured conditions of the Troodos mass, as a consequence of its uplift, facilitated deep weathering of the rocks, leading to the development of a smooth, mature topography, mantled with a thick cover of a diversity of soils (Robinson and Malpas, 1998). The soil cover of the central area is highly alkaline, while the soils on the slopes, lower down, are covered by neutral sheeted diabase. The weathering of the sedimentary rocks (chalks, marls, etc.) in the foothills that fringe Troodos, gave rise to alkaline, calcium–rich soils. None of these rock types belongs to the category of silica-oversaturated, which usually is associated with high Th and U elemental concentrations (Faure, 1986).

Since no systematic data on environmental radioactivity in Cyprus were available, a pilot project commenced in 2001 with the objective to systematically measure the terrestrial gamma radiation in the island, and determine its contribution to the annual effective dose equivalent to the population. There are two main features that make this study particularly important and interesting to radiometric studies. Firstly, it provides information on the geomorphological composition and the associated environmental radioactivity of such an area with a large variety of ophiolitic and sedimentary rock types cropping out over a relatively narrow area. Secondly, the analysis of selective elemental abundance (*Th/U*, *K/U*, and *K/Th* ratios) may also allow us to study the enrichment/depletion processes as a result of the complex metamorphic history, alteration and/or weathering that affected the investigating rocks (Chiozzi et al., 2002).

Tzortzis et al. (2003a) reported the first results of such measurements and the corresponding average annual effective dose rates equivalent to the population, using 28 soil samples collected from the main geological rock types appearing in the island. A more extensive study of terrestrial gamma radioactivity was performed after the collection of 115 soil samples from all over the island bedrock surface based on the different lithological units of the study area. Such a number of collected samples in combination with the relatively large area portion studied (about 6,000 $km^2$) can be considered that they widely cover the various outcropping geological formations. Other parts of the project aimed at measuring the concentration of radon in houses and public buildings (Anastasiou et al., 2003), as well as at detecting α−emitting radioisotopes by utilising radioanalytical techniques and high-resolution α-spectrometry (Pashalidis and Tsertos, 2003). In this paper, the results from this extensive survey regarding Th, U and K elemental concentrations in a wide variety of surface soils by means of high-resolution γ−ray spectrometry are presented. Further results from these investigations concerning activity concentrations and the associated dose rates due to terrestrial gamma radiation are going to be published elsewhere (Tzortzis et al., 2004). The experiments have been carried out in the Nuclear Physics Laboratory of the Department of Physics, University of Cyprus.

## 2. Materials and methods

### 2.1 Sample collection and preparation

A total of 115 surface soil samples has been collected throughout the whole accessible area[†] of the Republic of Cyprus (Fig. 1), which widely covered all the geological surface formations of the island. The official geological map of Cyprus (GSDC, 1995) was used to segregate the accessible island's area into seven different geological regions indicated in the simplified map shown in Fig. 1. Ten to twenty-eight soil samples were collected from different locations falling within the boundaries of each of the seven geological regions considered. Special emphasis was given to the Troodos ophiolitic complex, from which more than half (59) of the total number of samples was collected.

Soil samples were collected at 115 different sites with the only constraint that no sampling site should be taken close to a field boundary, a road, a tree, a building, or other obstruction. For each soil sample collected, an area of about 0.5 $m$ × 0.5 $m$ was marked and carefully cleared of debris to a few centimetres depth. Surface soils were then taken from different places randomly within the marked and cleared area, and mixed together thoroughly, in order to obtain a representative sample of that area. Each soil sample was labeled according to the geographical coordinates of the sampling area, and those coordinates were later used to indicate the position on the simplified map by an open circle point (see Fig. 1).

Collected soil samples were air-dried, sieved to remove stones and pebbles, and crushed to pass through a fine mesh sieve (~ 0.5 $mm$) to homogenise them. Finally, a part of each prepared soil sample was packed in a standard 1000-*mL* plastic Marinelli beaker that was hermetically sealed, dry-weighed and stored for about four weeks prior to counting. On the one hand, this led to samples in which the rather large soil mass used (being typically between 1.0 and 1.4 *kg*) was greatly homogenised within the beaker volume, and, on the other hand, ensured that equilibrium between $^{226}$Ra and $^{222}$Rn and their decay products was established. As far as the last point is concerned, one should notice that equilibrium is common in rocks older than $10^6$ *years*, and the $^{232}$Th series may be considered in equilibrium in most geological environments (Chiozzi et al., 2002).

---

[†] Samples from the northern part of Cyprus could not be collected because of the political status in the island since 1974 (see Fig. 1).

**2.2 Sample counting**

Soil sample containers were placed into the active volume of a shielded high-purity germanium (HPGe) detector with an efficiency of 33%, relative to a 3″×3″ NaI(Tl) scintillator, and measured for a counting time of 18 *hours*. A detailed description of the high-resolution γ–ray spectrometry system, as well as of the measurement and analysis techniques used in these investigations, has been presented elsewhere (Tzortzis et al., 2003a). The naturally occurring radionuclides considered in the present analysis of the measured γ–ray spectra are: $^{212}$Pb (with a main gamma energy at ~239 keV and a gamma yield of ~43.1%), $^{214}$Pb (~352 keV, ~37.1%), $^{214}$Bi (~609 keV, ~46.1%), $^{228}$Ac (~911 keV, ~29%), and $^{40}$K (~1461 keV, ~10.7%). Under the assumption that secular equilibrium was reached between $^{232}$Th and $^{238}$U and their decay products, the concentration of $^{232}$Th was determined from the average concentrations of $^{212}$Pb and $^{228}$Ac in the samples, and that of $^{238}$U was determined from the average concentrations of the $^{214}$Pb and $^{214}$Bi decay products (Hamby and Tynybekov, 2000; Tzortzis et al., 2003a). Thus, an estimation of radionuclide concentration of $^{232}$Th and $^{238}$U was obtained, whereas a direct measurement of $^{40}$K concentration was achieved. The environmental gamma-ray background at the laboratory site was determined using an empty Marinelli beaker under identical measurement conditions. From measurements prior, during, and after the experiments, it was found that the background levels in the laboratory were maintained constant during the whole period of the measurements. Background was then appropriately subtracted from the measured γ−ray spectrum of each sample.

The energy resolution (FWHM) achieved in the calibration measurements was 1.8 *keV* at the 1.33 *MeV* reference transition of $^{60}$Co. Depending on the background of the measured spectra, the Minimum Detectable Activity (MDA) was calculated to be $1.0 \times 10^{-2}$ *Bq kg$^{-1}$* for both $^{232}$Th and $^{238}$U, and $4.0 \times 10^{-2}$ *Bq kg$^{-1}$* for $^{40}$K, for the counting time of 18 *hours*. From these values, detection limits of $2.5 \times 10^{-3}$ *μg g$^{-1}$*, $8.1 \times 10^{-4}$ *μg g$^{-1}$*, and 1.3 *μg g$^{-1}$* were derived, for thorium, uranium and potassium elemental concentrations, respectively.

## 2.3 Calculation of elemental concentrations

Following the spectrum analysis, count rates for each detected photopeak and activity concentration (activity per unit soil mass) for each of the detected nuclides are calculated (Tzortzis et al., 2003a). Thus, for each sample studied the activity concentrations, $A_E$, of the $^{232}$Th, $^{238}$U and $^{40}$K radionuclides are determined in units of *Bq kg$^{-1}$*.

Activity concentrations of $^{232}$Th, $^{238}$U and $^{40}$K were then converted into elemental concentrations of thorium, uranium and potassium, respectively, according to the following expression:

$$F_E = \frac{M_E \cdot C}{\lambda_E \cdot N_A \cdot f_{A,E}} \cdot A_E \quad (1)$$

where $F_E$ is the fraction of element $E$ in the sample, $M_E$, $\lambda_E$, $f_{A,E}$ and $A_E$ are the atomic mass *(kg mol$^{-1}$)*, the decay constant *(s$^{-1}$)*, the fractional atomic abundance in nature and the measured activity concentration *(Bq kg$^{-1}$)*, respectively, of the corresponding radionuclide considered ($^{232}$Th, $^{238}$U, or $^{40}$K), $N_A$ is the Avogadro's number (6.023×10$^{23}$ *atoms mol$^{-1}$*), and $C$ is a constant with a value of 1,000,000 for Th and U or 100 for K. Hence, elemental concentrations are reported in units of *µg g$^{-1}$ (equivalent to ppm)* for thorium and uranium, and as a percentage *(%)* for potassium.

The uncertainty of the radioactivity measurements calculated, which is also applicable to the extracted Th, U, and K elemental concentration values, was typically in the range 3−10%. It has been calculated by taking into consideration the counting statistical error (~3%) and weighted systematic errors that mainly included the uncertainty in the efficiency calibration (0.5−8%) (Tzortzis et al., 2003b). However, additional uncertainties due to the dry weight estimation or sample homogeneity, as well as due to the assumption of secular equilibrium in the estimation of the $^{232}$Th and $^{238}$U activity, have not been taken into account.

## 3. Results and discussion

For the purpose of quality assurance, a control sample from a reference soil material (IAEA-326) was prepared in an identical 1000-mL Marinelli beaker (dry soil mass of 1.254 kg), and treated with respect to the measurement and analysis procedure as an unknown sample. The results obtained and the recommended values are summarised in Table 1. One may note that the recommended values (Bojanowski, 2001) are given in units of *Bq kg$^{-1}$* for the $^{232}$Th, $^{238}$U, and $^{40}$K activity concentrations, which are converted into corresponding elemental concentrations according to equation (1). As can be seen, the recommended mean values of the Th, U, and K elemental concentrations are reproduced by the measurements with a relative deviation of 5.4%, 8.0%, and 0.5%, respectively, demonstrating a rather good performance of the measurement and analysis technique utilised.

The calculated elemental concentrations of thorium, uranium and potassium for the 115 samples collected from all over the island bedrock surface based on the different lithological units of the study area are plotted in Figure 2. As can be seen, higher values of elemental concentrations are associated with formations of sedimentary origin (regions R1, R2, R3 and R5) and lower values are associated with formations belonging to the Troodos ophiolitic complex (regions R4, R6 and R7). In particular, region R4, which covers the central part of the Troodos massif that contains basic and ultrabasic plutonic rocks (see legend in Fig. 1), appears to present the lowest concentrations in Th, U and K.

Considering all samples studied, elemental concentrations of thorium ranged from $2.5\times10^{-3}$ *µg g$^{-1}$* to 9.8 *µg g$^{-1}$*, from $8.1\times10^{-4}$ *µg g$^{-1}$* to 3.2 *µg g$^{-1}$* for uranium, and from $1.3\times10^{-4}$ *%* to 1.9 *%* for potassium. It should be noted that the lower values given correspond to the detection limit derived for each of the three elements, since some samples (mainly in the region R4) exhibited nearly zero net activity concentration after the subtraction of the ambient background in the laboratory site. Arithmetic mean values of measured elemental concentrations over all samples are: (1.2 ± 1.7) *µg g$^{-1}$*, (0.6 ± 0.7) *µg g$^{-1}$* and (0.4 ± 0.3) *%,* for thorium, uranium and potassium, respectively, while median values obtained worldwide are 7.4 *µg g$^{-1}$*, 2.8 *µg g$^{-1}$*, and 1.3 *%,* respectively. The latter mean values are derived by transforming the corresponding worldwide average activity concentrations of 30, 35, and 400 *Bq kg$^{-1}$* (UNSCEAR 2000 Report) for $^{232}$Th, $^{238}$U and $^{40}$K radionuclides, into Th, U, and K elemental concentrations, respectively, using equation (1). This result reveals that the mean concentrations of the radioactive elements of Th, U and K, as obtained from all the geological formations studied in Cyprus, are by a factor of three to six lower than the corresponding values reported worldwide. Region 5, which is mainly composed of calcarenites and sandstones, exhibits the highest concentration in Th and K with mean values of (4.1 ± 2.3) $\mu g\ g^{-1}$ and (0.6 ± 0.3) %, respectively, while Region 2, mainly composed of chalks, marls and gypsum, reveals the highest concentration in U with a mean of (1.8 ± 0.8) $\mu g\ g^{-1}$. Although comparable and within the range of some worldwide areas, the above presented highest mean elemental concentrations of Th, U and K are still below the worldwide mean concentrations derived from the corresponding activity concentrations reported in the UNSCEAR 2000 Report.

For a more general and representative overview, arithmetic mean, minimum, and maximum concentrations for each one of the seven geological regions are summarised in Table 2. The highest concentrations of Th, U and K were measured in soil samples belonging to sedimentary formations, whereas the corresponding mean concentrations associated with soils that stem from the Troodos ophiolitic complex are significantly lower. In Figure 3, the data for the arithmetic mean concentrations of Th, U and K for each geological formation studied are illustrated.

The results presented in this paper are in general agreement with the first measurements of activity and elemental concentrations of naturally occurring radionuclides in 28 samples from characteristic geological rock types of Cyprus reported by Tzortzis et al. (2003a). In that study, soil samples of sedimentary composition appear again with higher mean elemental concentrations in the radioisotopes of interest than those exhibited by soils of ophiolitic origin as following: (4.1 ± 0.9) $\mu g\ g^{-1}$ against (0.5 ± 0.1) $\mu g\ g^{-1}$ for thorium, (1.7 ± 0.4) $\mu g\ g^{-1}$ against (0.5 ± 0.4) *µg g*$^{-1}$ for uranium, and (0.7 ± 0.1) *%* against (0.5 ± 0.2) *%* for potassium, for samples of sedimentary and ophiolitic origin, respectively. Arithmetic mean values of the measured elemental concentrations over all samples were found to be (2.8 ± 0.7) *µg g*$^{-1}$, (1.3 ± 0.3) *µg g*$^{-1}$ and (0.6 ± 0.1) *%,* for thorium, uranium and potassium, respectively. For a direct comparison, one should notice that in the present study more than half (~51%) of the samples collected are of ophiolitic origin, which reveal very low activity concentration levels compared to those of sedimentary origin. The corresponding percentage contribution of samples of ophiolitic origin in the first study (Tzortzis et al., 2003a) was ~39%, and this might explain why the results of the activity concentration and dose rates are slightly higher compared to the corresponding values of the present study.

In Table 3, summary results on elemental concentrations, derived from similar investigations conducted in other countries worldwide with emphasis on regions close and around the Mediterranean Sea, are presented. As can be seen, even the maximum elemental concentration values obtained from this study fall within the lowest range of most reported values from other worldwide areas. In other words, since natural radioactivity is directly related to the content in Th, U, and K radioactive elements of the rock from which the soils originate, the island of Cyprus can be considered as one of the world areas that exhibit very low levels of natural radioactivity.

Verdoya at al. (2001) state that the original Th, U, and K concentrations in rocks may vary because of alteration or metamorphic processes. Figure 4 displays the *Th/U*, *K/U*, and *K/Th* ratios, which may provide an indication whether relative depletion or enrichment of radioisotopes had occurred. The best-fitting relation between Th and K versus U is of linear type, with a correlation coefficient of 0.93 and 0.84, respectively. The theoretically expected *Th/U* ratio for normal continental crust is about 3.0, while

**11**

the corresponding value obtained for this ratio is (2.0 ± 0.4), which is still close to the expected value. Chiozzi et al. (2002) comment on this fact that it seems to indicate that no significant fractionation during weathering or involvement in metasomatic activity of the radioelements had occurred. The *K/U* ratio has been calculated to be equal to $(2.8 \pm 0.8) \times 10^3$, but it should be stated that this value is highly variable (e.g. Roger and Adams, 1969). In the correlation between K and Th, the best-fitting relation is again of linear type with a correlation coefficient of 0.90. The slope in this case is $(1.4 \pm 0.3) \times 10^3$, which might be considered as close to the value of $2.5 \times 10^3$ reported by Chiozzi et al. (2002), and also close to the typical value obtained in a large variety of unaltered rock formations by Galbraith and Saunders (1983). Finally, it is worth pointing out in this context that by a closer inspection of the *K/U* and *K/Th* data shown in Fig. 4, one recognises that the soils stemming from the Troodos ophiolite complex (R4, R6, R7 regions) indicate rather larger slopes as compared to those of sedimentary origin (R1, R2, R3, R5). This might reflect a quite different enrichment/depletion process of the two main soil categories classified, although it is widely believed that the soils originated from the sediments have been developed from the underlying main ophiolithic complex. It is clear that before definite conclusions on this issue are drawn, more systematic data and further analysis is required.

## 4. Conclusions

High-resolution γ–ray spectrometry was used to determine elemental concentrations of the radioactive elements of thorium, uranium, and potassium in a number of 115 surface soil samples collected from all over the island of Cyprus. The number and the surface distribution of the samples collected can be considered as widely and sufficiently covering the various geological formation areas. Soils originated from the Troodos ophiolitic complex appear generally to have lower naturally occurring radionuclide concentrations, compared to those of sedimentary origin. In the present study, arithmetic mean concentrations of 1.2 $\mu g\ g^{-1}$, 0.6 $\mu g\ g^{-1}$, and of 0.4 % for thorium, uranium and potassium radioactive elements, respectively, are derived. These values fall within the lowest range of those measured at worldwide scale by other authors and, more specifically, are by a factor of three to six lower than the reported world average values of 7.4 $\mu g\ g^{-1}$ (Th), 2.8 $\mu g\ g^{-1}$ (U) and 1.3% (K) in the UNSCEAR 2000 Report.

The results obtained from the present wide-range investigations are in general agreement with the first studies on activity and elemental concentrations of naturally occurring radionuclides and on their associated dose rates in various rock types (Tzortzis et al., 2003a). They further confirm that the dose rates due to naturally occurring radiation in Cyprus are significantly lower (by a factor of five on an average) than the corresponding radioactivity levels reported from other areas worldwide.

## Acknowledgements

This work is financially supported by the Cyprus Research Promotion Foundation (Grant No. 45/2001), and partially by the University of Cyprus.

**TABLE CAPTIONS**

**Table 1.** Quality control measurements using the reference soil material IAEA-326 that has been treated as an "unknown" sample. The recommended gamma activity concentration values ($Bq\ kg^{-1}$) have been converted into elemental concentrations according to equation (1).

| **Reference soil material IAEA-326** | | | |
|---|---|---|---|
| (Dry soil mass used: 1.254 $kg$) | | | |
| | Recommended elemental concentrations *(Bojanowski 2001)* | | Measurement results[**] |
| **Radioelement** | **Mean value** | **95% Confidence interval** | **Concentration ± S.D.** |
| **Th** ($\mu g\ g^{-1}$) | *9.71* | *9.26 – 10.15* | *10.23 ± 0.15* |
| **U** ($\mu g\ g^{-1}$) | *2.38* | *2.28 – 2.49* | *2.57 ± 0.05* |
| **K** *(%)* | *1.91* | *1.88 – 1.94* | *1.90 ± 0.02* |

[**] Measurements were taken at three times on the reference soil.

**Table 2.** Arithmetic Mean (A.M.), Standard Deviation (S.D.), Minimum (Min), and Maximum (Max) elemental concentrations of thorium, uranium, and potassium in soil samples from the main geological regions studied (see the legend in Fig. 1).

| | Region | Number of samples | Average elemental concentration | | | | | |
|---|---|---|---|---|---|---|---|---|
| | | | Th ($\mu g\ g^{-1}$) | | U ($\mu g\ g^{-1}$) | | K (%) | |
| | | | A.M./ S.D. | Min/ Max | A.M./ S.D. | Min/ Max | A.M./ S.D. | Min/ Max |
| Sedimentary Rocks | R1 | 23 | *1.4E+0* | *9.8E-2* | *7.8E-1* | *6.9E-2* | *4.7E-1* | *4.7E-2* |
| | | | *2.0E+0* | *9.8E+0* | *5.8E-1* | *2.8E+0* | *4.2E-1* | *1.9E+0* |
| | R2 | 14 | *2.7E+0* | *1.3E+0* | *1.8E+0* | *7.1E-1* | *5.0E-1* | *2.9E-1* |
| | | | *1.5E+0* | *7.1E+0* | *8.2E-1* | *3.2E+0* | *2.2E-1* | *1.0E+0* |
| | R3 | 10 | *2.1E+0* | *8.5E-1* | *8.2E-1* | *3.6E-1* | *3.9E-1* | *1.6E-1* |
| | | | *8.5E-1* | *3.2E+0* | *3.5E-1* | *1.3E+0* | *1.5E-1* | *6.1E-1* |
| | R5 | 9 | *4.1E+0* | *1.3E+0* | *9.7E-1* | *5.6E-1* | *6.3E-1* | *2.6E-1* |
| | | | *2.3E+0* | *7.0E+0* | *3.4E-1* | *1.6E+0* | *3.1E-1* | *1.0E+0* |
| | **R1-R5** | 56 | *2.3E+0* | *9.8E-2* | *1.1E+0* | *6.9E-2* | *4.8E-1* | *4.7E-2* |
| | | | *2.0E+0* | *9.8E+0* | *7.1E-1* | *2.8E+0* | *3.2E-1* | *1.9E+0* |
| Ophiolitic Rocks | R4 | 28 | *8.0E-2* | *2.5E-3* | *3.7E-2* | *8.1E-4* | *9.2E-2* | *1.3E-4* |
| | | | *1.8E-1* | *9.0E-1* | *4.3E-2* | *1.4E-1* | *9.0E-2* | *3.0E-1* |
| | R6 | 21 | *3.6E-1* | *2.5E-3* | *1.4E-1* | *9.7E-3* | *2.3E-1* | *3.7E-2* |
| | | | *4.0E-1* | *2.0E+0* | *9.0E-2* | *4.5E-1* | *1.5E-1* | *7.4E-1* |
| | R7 | 10 | *4.4E-1* | *1.7E-1* | *2.0E-1* | *7.6E-2* | *5.2E-1* | *5.4E-2* |
| | | | *3.5E-1* | *1.4E+0* | *1.4E-1* | *5.5E-1* | *3.7E-1* | *1.0E+0* |
| | **R4-R7** | 59 | *2.4E-1* | *2.5E-3* | *1.0E-1* | *8.1E-4* | *2.1E-1* | *1.3E-4* |
| | | | *3.4E-1* | *2.0E+0* | *1.0E-1* | *5.5E-1* | *2.4E-1* | *1.0E+0* |
| | **R1-R7** | 115 | *1.2E+0* | *2.5E-3* | *5.7E-1* | *8.1E-4* | *3.5E-1* | *1.3E-4* |
| | | | *1.7E+0* | *9.8E+0* | *7.0E-1* | *3.2E+0* | *3.1E-1* | *1.9E+0* |

**Table 3.** Summary of Th, U, and K elemental concentrations derived from reported values of gamma activity concentrations ($Bq\ kg^{-1}$) in soil samples from work conducted worldwide.

| Region | Th ($\mu g\ g^{-1}$) | U ($\mu g\ g^{-1}$) | K (%) | Reference |
|---|---|---|---|---|
| United States | 8.9 ± 4.2 | 3.0 ± 2.5 | – | Myrick et al. (1983) |
| Alps-Apennines, Italy | 0.3 – 16.7 | 0.3 – 5.6 | 0.14 – 5.14 | Chiozzi et al. (2002) |
| Amman, Jordan | 7.1 | 4.6 | 1.7 | Ahmad et al. (1997) |
| Karak, Jordan | 6.7 | 18.5 | 1.4 | Ahmad et al. (1997) |
| Instanbul, Turkey | 9.1 | 1.7 | 1.1 | Karahan and Bayulken (2000) |
| Costal area, Aegean sea, Greece | 17.5 ± 6.2 | 7.5 ± 3.8 | 2.9 ± 1.2 | Florou and Kritidis (1992) |
| Taiwan | 10.8 | 2.4 | 1.4 | Yu-Ming et al. (1987) |
| Russaifa, Jordan | 2.1 – 6.7 | 3.9 – 42.4 | 0.1 – 1.0 | Al-Jundi (2002) |
| Italy | 18 – 21 | 4.6 – 5.7 | 1.9 – 2.5 | Bellia et al. (1997) |
| Spain | 3.2 – 20.9 | 1.6 – 57.6 | 1.0 – 2.3 | Martinez-Aguirre and Garcia-Leon (1997) |
| Canary islands | 13.3 | 3.6 | 2.2 | Fernandez et al. (1992) |
| Spain | 1.7 – 50.3 | 1.1 – 13.4 | 0.2 – 5.2 | Baeza et al. (1992) |
| Rajasthan, India | 10.6 – 26.1 | 2.4 – 6.3 | 0.2 – 0.5 | Nageswara et al. (1996) |
| Cyprus | <1 – 9.8 | <1 – 3.2 | <1 – 1.9 | Present study |
| Worldwide average | 7.4 | 2.8 | 1.3 | UNSCEAR report (2000) |

**FIGURE CAPTIONS**

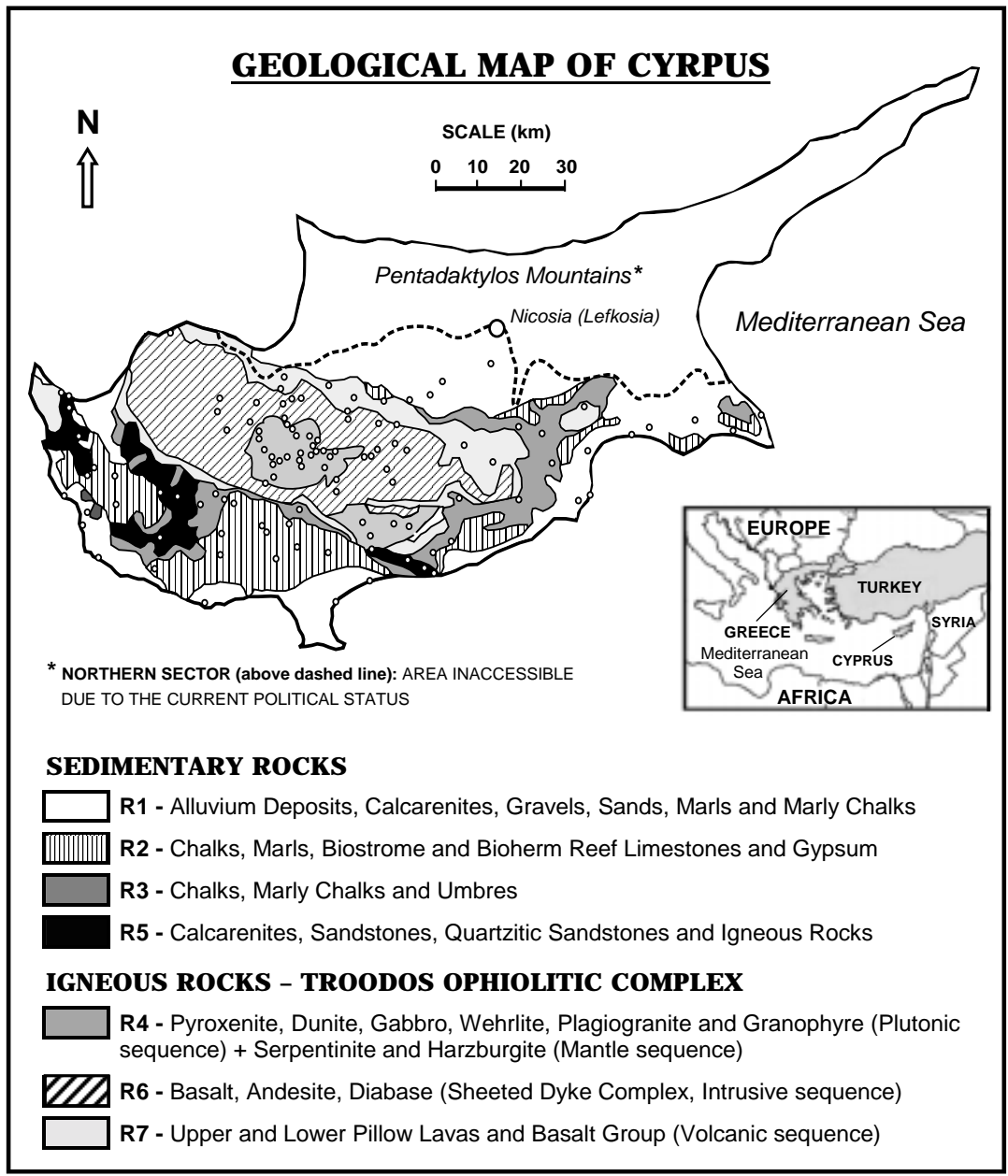

**Figure 1**. Simplified geological map of Cyprus, indicating the seven geological regions examined in this survey; the various rock formations are described in the legend. Open circle points indicate the locations from where soil samples were collected in each geological region.

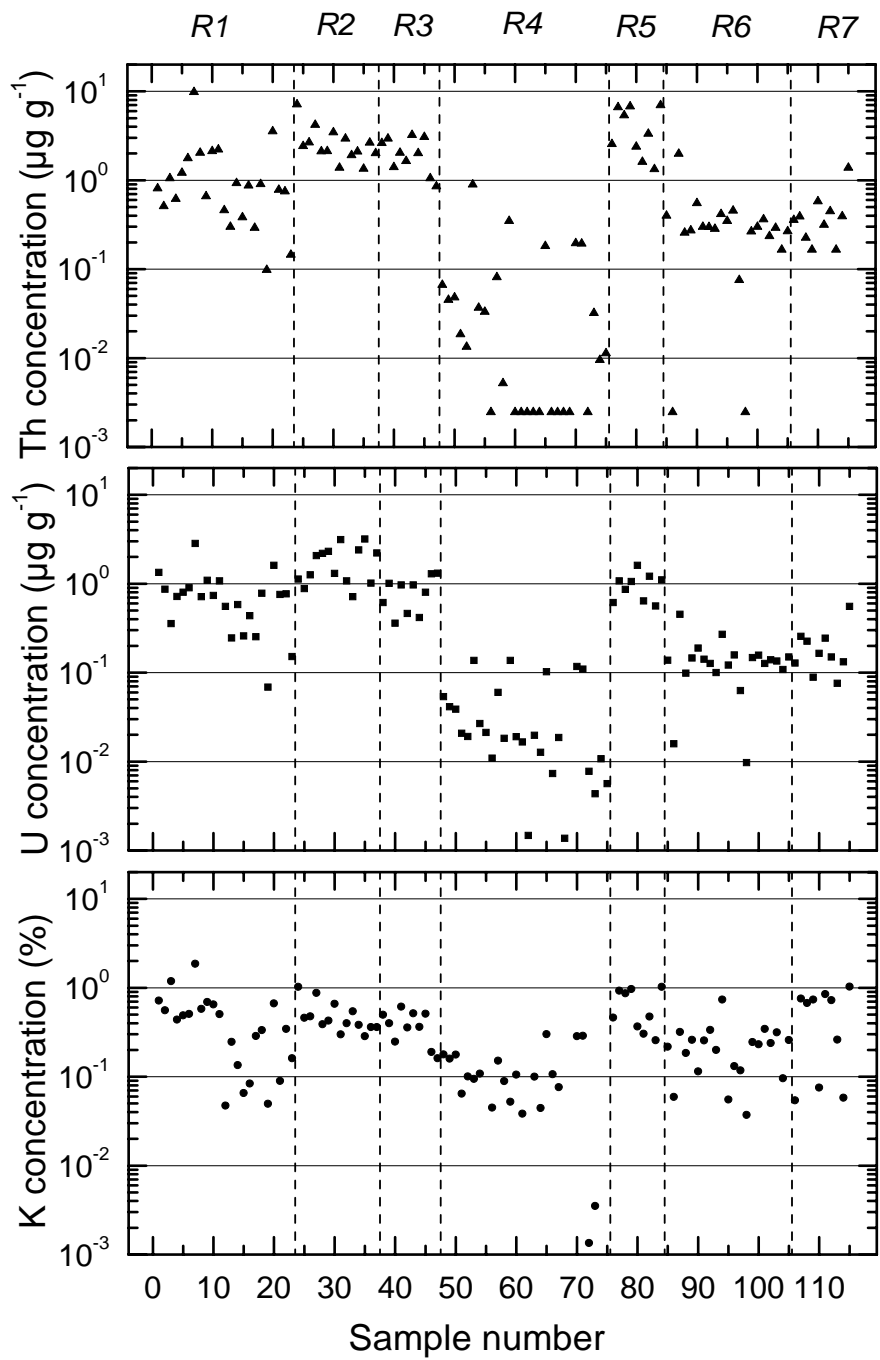

**Figure 2.** Thorium, uranium and potassium elemental concentrations from all the soil samples studied. The vertical dashed lines correspond to the borders of each of the seven regions considered (see Fig. 1). In the region R4, values lower than $1 \times 10^{-3}$ are not shown in the picture.

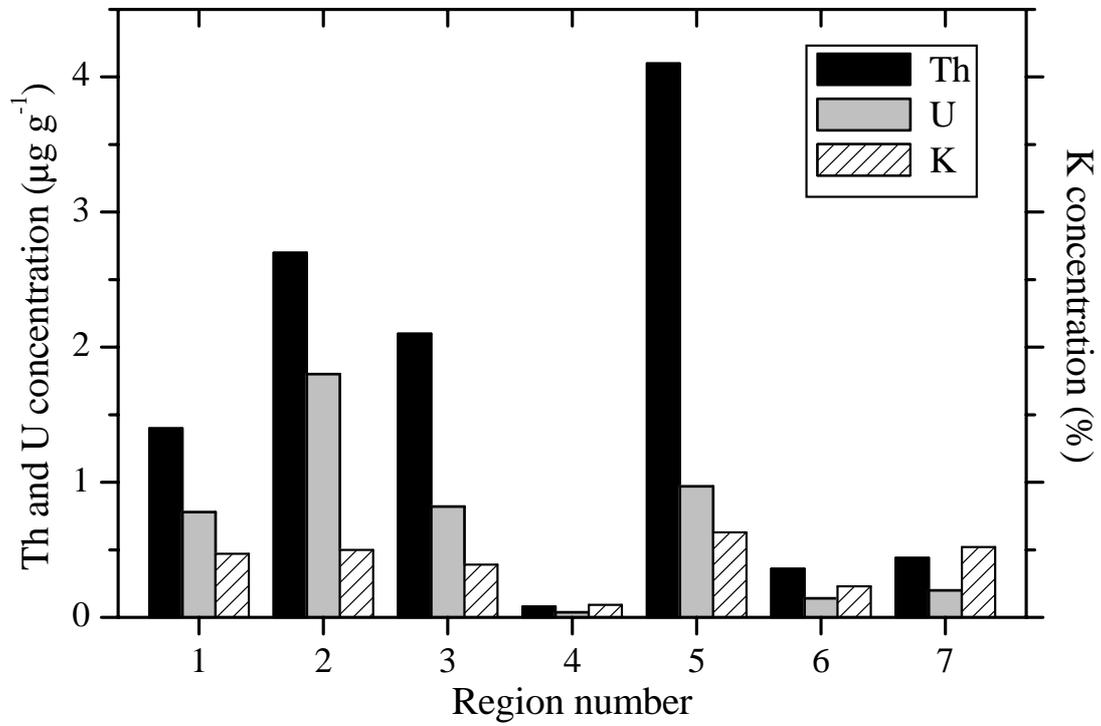

**Figure 3.** Arithmetic mean values of thorium, uranium and potassium elemental concentrations for the seven main geological regions from which the soil samples were collected.

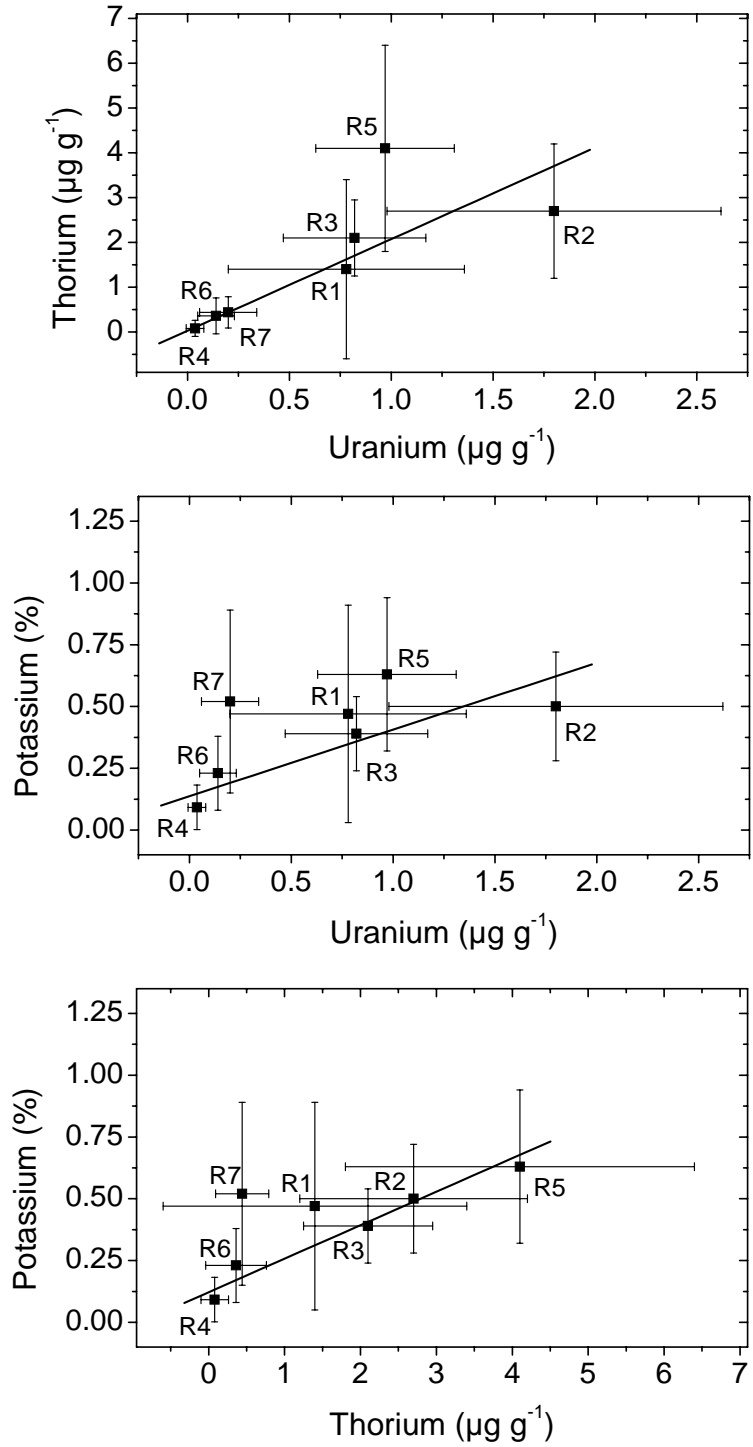

**Figure 4.** Th versus U, K versus U, and K versus Th arithmetic mean concentrations and standard deviations (error bars) of geological regions of Table 2, grouped according to their geological composition illustrated in the legend of Fig. 1. The solid straight lines represent the best fitting relation to the data points.